\documentclass[aps,nofootinbib,twocolumn,prd,eqsecnum,showpacs,showkeys,preprintnumbers]{revtex4-1}
\usepackage[caption=false]{subfig}
\usepackage{graphicx}
\usepackage{amsmath}
\usepackage{amsfonts}
\usepackage{amssymb}
\usepackage{color}
\usepackage{bm}
\usepackage{mathrsfs}
\usepackage{epstopdf}
\usepackage{url}
\usepackage{footnote}
\usepackage{textcomp}
\usepackage{ulem}
\usepackage{esint}
\usepackage[unicode=true, pdfusetitle,
 bookmarks=true,bookmarksnumbered=false,bookmarksopen=false,
 breaklinks=false,pdfborder={0 0 1},backref=false,colorlinks=false]{hyperref}
\usepackage{multirow}
\usepackage{pifont}

\begin{document}
\title{Dynamical system approach of non-minimal coupling in AdS/CFT cosmology}

\author{Aatifa Bargach}
\email{a.bargach@ump.ac.ma}
\author{Farida Bargach}
\email{f.bargach@ump.ac.ma}
\author{Taoufik Ouali}
\email{ouali\_ta@yahoo.fr}
\affiliation{Laboratory of Physics of Matter and Radiation, \\
University of Mohammed first, BP 717, Oujda, Morocco}

\date{\today }

\begin{abstract}
We study the dynamical system approach of non minimally coupled scalar field to induced gravity on the brane in the framework of the AdS/CFT correspondence. In this context, we derive the modified Friedmann equation and the equation of motion. The dynamics of this model are studied by rewriting the cosmological field equations in the form of a system of autonomous differential equations. In particular, the analysis is considered by investigating an exponential potential and a monomial form of the non minimal coupling function. We show that, for sufficient conditions, a past de Sitter attractor solution is obtained in the case of a minimal coupling, meanwhile a future de Sitter attractor solutions  is obtained 
for a conformal coupling and for a non minimal coupling.

\end{abstract}
\maketitle
\section{Introduction}
In recent years, extra dimensional space-time has taken considerable research interest such as braneworld models in which ordinary matter fields are considered to live on the boundary of a high-dimensional bulk space-time, the brane. In particular, Randall and Sundrum \cite{Randall:1999vf} proposed a model (RSII) in which a brane with a positive tension is embedded in five dimensional anti de Sitter space. The cosmology of this model shows that at low energies general relativity on the brane can be recovered, while in high energy limit gravity becomes five dimensional. The RSII
model can also be  considered as an example of the holographic principle \cite{adsholo1, adsholo2, adsholo3, adsholo4} that has emerged in M theory. Indeed, holography suggests that higher-dimensional gravitational dynamics may be determined from knowledge of the fields on a lower-dimensional boundary. A concrete illustration of this holographic principle is The AdS/CFT correspondence. This kind of correspondence asserts that there is an equivalence between a gravitational theory in d-dimensional anti de Sitter space-time and a conformal field theory living in a ($d-1$)-dimensional boundary space-time \cite{intro2}. This equivalence or duality is best understood in the context of string theory with d=5, where the duality relates type IIB superstring theory on $AdS_{5} \times S_{5}$, and $N=4$ supersymmetric Yang Mills theory with gauge group SU(N) in four dimensions \cite{intro3, intro4}. The RSII model with its $AdS_{5}$ metric satisfies this correspondence to lowest perturbative order \cite{RSAds}. In this paper, the AdS/CFT correspondence is the subject of our framework in order to derive the modified equations.\par

On the other hand, Dvali, Gabadadze and Porrati \cite{Dvali:2000hr} suggested a model with a bulk as a flat Minkowski spacetime, but a reduced gravity term appears on the brane without tension. This setup is based on a modification of the gravitational theory in an induced gravity perspective \cite{inducedper1, inducedper2, inducedper3, inducedper4}. Generally, induced gravity (IG) effect can be viewed as a quantum correction in any braneworld model for instance the Randall-Sundrum model.
{The cosmology of IG corrections to RS models have been treated by many authors \cite{Kofinas:2001es, Deffayet:2000uy, Kiritsis:2002ca, Maeda:2003ar, Papantonopoulos:2004bm, Zhang:2004in, BouhmadiLopez:2004ax}. In the spirit of IG corrections, one can consider a non-minimal coupling (NMC) of the scalar field to the intrinsic curvature on the brane. Scalar fields arise in a natural way in particle physics and act as a candidate for both models of the early universe and late-time acceleration. Trapped on a brane, scalar fields provide a simple dynamical model for matter fields and have been investigated widely in the literature \cite{Maeda:2000mf, GonzalezDiaz:2000ad, Majumdar:2001mm, Nunes:2002wz, Sami:2004ic}.
The motivation for including a NMC term arises at the quantum level when quantum corrections to the scalar
field theory are considered. This kind of NMC to gravity has been discussed enough  in four dimensions \cite{Futamase:1987ua, Salopek:1988qh, Fakir:1990eg, Amendola:1990nn, Kaiser:1994vs, Bezrukov:2007ep, Bauer:2008zj, Park:2008hz, Linde:2011nh, Kallosh:2013maa, Kallosh:2013tua, Chiba:2014sva, Boubekeur:2015xza, Pieroni:2015cma, Salvio:2017xul},  in extra dimensions \cite{mariam, Nozari:2012cy, Bogdanos:2006qw, Farakos:2006sr} and also in IG \cite{Accetta:1985du}. It turns out that in general relativity, the coupling constant is valued to $1/6$ \cite{deals}. We analyze this coupling value as a particular case.\par}
{
This work is also a complementary of the minimally coupled scalar field to the gravity \cite{Quiros:2008hv} where the authors considered the dynamics of a scalar field in a Dvali-Gabadadze-Porrati brane, of  those of  \cite{Gonzalez:2008wa, Escobar:2012cq}   where the scalar field is trapped in the RSII brane and of the IG of a NMC scalar field \cite{mariam, Nozari:2012cy}. Both of above the works used the geometrical approach to derive the effective Einstein equations projected onto the brane. Here, we use the AdS/CFT correspondence to derive the modified equations and investigate the effects of such modifications to the dynamic behavior of the universe where the scalar field is localized in the brane.  \par}

{
Since explicit solutions of the evolution equations cannot be obtained in this setup, the theory of dynamical systems has proven to be a very powerful scheme to obtain exact solutions and a qualitative description of the global dynamics. Dynamical systems methods are widely used in cosmology and have been applied to  extended theories of gravity  \cite{ds1, ds2, ds3, Odintsov:2015wwp, Odintsov:2017icc, Odintsov:2017tbc}, to braneworld theories \cite{Dutta:2015jaq, Escobar:2013js, Gonzalez:2008wa, Escobar:2011cz, Zonunmawia:2018xvf} and to a NMC scalar fields \cite{Sakstein:2015jca, Hrycyna:2015eta, Bhattacharya:2015wlz}. This motivates us to consider these techniques in order to obtain exact solutions and a qualitative description of the global dynamics of our model.\\

In the present paper, we study the impact of the modifications of the Friedmann equation of non minimally coupled scalar field to induced gravity on the brane in the framework of the AdS/CFT correspondence using the DSA.
We apply the DSA to study a specific potential and a NMC function choices, namely an exponential potential and a monomial form of NMC function. By applying dynamical system methods, we focus on the early and late time attractors behavior of the state of the universe.} \\

{This paper is organized as follows. In Sec. \ref{sec2}, we present the model and then use the AdS/CFT correspondence to give the effective Einstein equations projected on the brane. In Sec. \ref{sec3}, we present the basic cosmological equations to describe the evolution of a NMC scalar field such as the modified Friedmann equation and the equation of motion. In Sec. \ref{sec4}, a DSA is developed by fixing the choice of an exponential potential and a monomial form of the coupling function. The stability behavior of critical points is examined using linear stability analysis and when necessary center manifold theory as well as numerical perturbation techniques. Using the DSA we show that this model present a de Sitter brane as attractor solutions. We present our summary and conclusions in Sec. \ref{conclusion}. Finally, in appendix A we apply the center manifold theory to study the stability properties of the non hyperbolic critical point.}\\

\section{The setup}\label{sec2}
In this work, we will analyse the model described by the action \cite{mariam, Nozari:2012cy}
\begin{multline}\label{action}
S=\int_{bulk}d^{5}x\sqrt{-g^{\left( 5\right) }}\left( \frac{1}{2\kappa_{5}^{2}}R_{5}-\Lambda
_{5}\right)\\
+\int_{brane}d^{4}x\sqrt{-\overset{}{g}}%
\left(-\Lambda_{4}+ \mathfrak{L}_{\phi }\right),
\end{multline}
where $\kappa_{5}^{2}$ is the 5D gravitational constant, $R_{5}$ is the Ricci scalar of the five-dimensional metric $g^{(5)}$ and $\Lambda_{5}$ is the bulk cosmological constant. $g$ is the induced metric on the brane, $\Lambda_{4}$ is the brane tension and $ \mathfrak{L}_{\phi }$, the Lagrangian density of the non minimal scalar field localized on the brane, is defined as
\begin{equation}
\mathfrak{L}_{\phi }= f(\phi ) R+\frac{1}{2}%
g^{\mu \nu }\nabla _{\mu }\phi \nabla _{\upsilon }\phi -V(\phi ).
\end{equation}
where $\nabla _{\mu }$ is the covariant derivative associated with the induced metric on the brane, $V(\phi )$ is the scalar field potential, and $f(\phi )\equiv \frac{1}{2} \left(\frac{1}{\kappa_{4}^{2}}-\alpha(\phi)\right)$ is a coupling between the scalar field $\phi$ and the induced gravity $R$.\par
The gravitational field equations through the AdS/CFT correspondence are obtained by extremizing the variation of the dual action with respect to the metric tensor \cite{jcap, boer,lidsey}
\begin{equation}\label{Gmod}
  \frac{1}{8\pi G_{N}} G^{(4)}_{\mu\nu}= T_{\mu\nu} + T_{\mu\nu}^{CFT}- \lambda g_{\mu\nu}.
\end{equation}
where $G_{N}$ denotes the Newton's constant ($8\pi G_{N}= \kappa_{4}^{2}$), $T_{\mu\nu}$ is the total energy-momentum tensor and $T_{\mu\nu}^{CFT}$\footnote{$T_{\mu\nu}^{CFT}$ represents the term $V_{\mu\nu}$ in Refs. \cite{lidsey, jcap}.} denotes the CFT energy momentum tensor.\\
The total energy-momentum tensor $T_{\mu\nu}$ and the effective cosmological constant on the brane $\lambda$ are given by \cite{mariam}
\begin{equation}\label{Tmunu}
  T_{\mu\nu} = T_{\mu\nu}^{(\phi)}+ T_{\mu\nu}^{(f)}-2 f(\phi) G_{\mu\nu},
\end{equation}
\begin{equation}\label{lambda4}
  \lambda = \frac{\kappa^{2}}{2} \Lambda_{5}+ \frac{\kappa^{4}}{12}\Lambda_{4}^{2}.
\end{equation}
The energy-momentum tensor of the conformal field theory, $ T_{\mu\nu}^{ \;CFT}$, cannot be written in the local covariant form, however its trace writes  \cite{lidsey}
\begin{equation}\label{tcft}
  T_{\mu}^{CFT \;\mu} = c \left(R^{\alpha}_{\;\; \beta}R^{\beta}_{\;\; \alpha}- \frac{1}{3} R^{2}\right),
\end{equation}
where $c$ is the conformal anomaly related to the AdS/CFT length.\\
The total energy-momentum tensor Eq. (\ref{Tmunu}) has been split of into a scalar field energy-momentum tensor,
\begin{equation}\label{tmunuphi}
T_{\mu\nu}^{(\phi)} = \nabla_{\mu}\phi \nabla_{\nu} \phi - \frac{1}{2} g_{\mu\nu} (\nabla\phi)^{2} - g_{\mu\nu} V(\phi),
\end{equation}
and into a non-minimal coupling energy momentum tensor
\begin{equation}\label{tmunuf}
  T_{\mu\nu}^{(f)} =  2 \nabla_{\mu}\nabla_{\nu} f(\phi) - 2 \Box f(\phi) g_{\mu\nu}.
\end{equation}
For a spatially flat Friedmann-Robertson-Walker universe (FRW), we may define the conformal field energy momentum tensor as \cite{jcap, lidsey}
 \begin{equation}\label{cft}
   T^{\mu\; CFT}_{\nu}\equiv \begin{pmatrix} -\sigma & \sigma \\ 0 & \sigma_{p} \delta^{i}_{j}\end{pmatrix}.
 \end{equation}
The Bianchi identity, $\nabla^{\mu}G_{\mu \nu} = 0$, and the equation of conservation of the energy-momentum, $\nabla^{\mu}T_{\mu \nu} = 0$, implies that $\nabla^{\mu} T^{CFT}_{\mu \nu} = 0 $, which amounts to
\begin{equation}\label{conservation}
  \dot{\sigma}+ 3 H (\sigma+\sigma_{p})=0,
\end{equation}
where $H$ is the Hubble parameter.\par
Furthermore, the trace of the conformal anomaly equation (\ref{tcft}) simplifies to
\begin{equation}\label{trace}
  \sigma - 3 \sigma_{p} = 24 c H^{2} (\dot{H}+H^{2}),
\end{equation}
and Eq. (\ref{conservation}) becomes
\begin{equation}\label{sigma}
  \dot{\sigma}+4 H \sigma -24 c H^{3}(H^{2}+ \dot{H})=0,
\end{equation}
whose solution reads
 \begin{equation}\label{sigma solution}
   \sigma = \chi_{rad}+ 6 cH^{4},
 \end{equation}
where $\chi_{rad}$ is an effective radiation term. During inflation, this term is rapidly redshifted as $a^{-4}$ away and its contribution can be neglected \cite{lidsey}.\\
\section{BASIC COSMOLOGICAL EQUATIONS}\label{sec3}
 In this section we will consider the following spatially flat isotropic and homogeneous FRW brane
 \begin{equation}\label{FRW}
   ds^{2}=- dt^{2}+a(t) \delta_{ij} dx^{i}dx^{j},
 \end{equation}
 where $a(t)$ is the scale factor, $\delta_{ij}$ is a symmetric $3-$dimentional metric and $x^{i}$, $i = 1, 2, 3$ are the comoving spatial coordinates.\\
From the $(00)$-component of the field equations (\ref{Gmod}) together with the equations (\ref{Tmunu}) and (\ref{sigma solution}), the modified Friedmann equation on this spatially flat brane can be obtained as
\begin{equation}\label{freidman}
  H^{2}= \frac{\kappa_{eff}^{2}}{3}\left(\rho + \lambda + 6 c H^{4} \right),
\end{equation}
where $\rho$ is the total energy density and the effective gravitational coupling, $\kappa^{2}_{eff}$, is given by
\begin{equation}\label{mphi}
   \kappa_{eff}^{2} = \frac{\kappa^{2}_{4}}{2-\kappa^{2}_{4}\alpha(\phi)}.
\end{equation}
Following the notation introduced in \cite{mariam}, we can write the total energy density and the pressure of the universe respectively as
\begin{equation}\label{rho}
  \rho = \rho^{(\phi)}+ \rho^{(\alpha)},\qquad p = p^{(\phi)}+ p^{(\alpha)},
\end{equation}
 where
\begin{equation}\label{rhophi}
\rho^{(\phi)}= \frac{1}{2} \dot{\phi}^{2}+V(\phi),\qquad p^{(\phi)} =  \frac{1}{2} \dot{\phi}^{2}-V(\phi),
\end{equation}
\begin{equation}\label{rhof}
\rho^{(\alpha)}= 3 H \frac{d\alpha(\phi)}{dt}\quad\text{and}\quad p^{(\alpha)}= -2 H \frac{d\alpha(\phi)}{dt}- \frac{d^{2}\alpha(\phi)}{dt^{2}}.
\end{equation}

The modified Friedmann equation Eq.(\ref{freidman}) can be rewritten as
\begin{equation}\label{freidmanmod}
 H^{2} = \frac{1}{4 c \kappa_{eff}^{2}} \left [ 1\pm \sqrt{1- \frac{(\rho+\lambda)}{\rho_{max}}}\right],
\end{equation}
where $\rho_{max} = \frac{3}{8c \kappa_{eff}^{4}}$.\\
In the limit $\alpha(\phi) \rightarrow 0$ we recover the modified Freidmann equation of the Randall-Sundrum cosmology in the context of the AdS/CFT correspondence \cite{jcap,lidsey} with a minimally coupled scalar field.\par
Furthermore, the modified Raychaudhuri equation can be deduced from Eq. (\ref{Gmod}) as
\begin{equation}\label{Ray}
  \frac{\ddot{a}}{a} = - \frac{\kappa_{eff}^{2}}{6} \left( \rho+ 3 p-2 \lambda - 12 c H^{2} (2 \frac{\ddot{a}}{a} -H^{2})\right).
\end{equation}
Finally, minimising the action \eqref{action} with respect to variation of the scalar field, $\phi$, we obtain the equation of motion in the FRW geometry as
\begin{equation}\label{feild}
  \ddot{\phi}+3H\dot{\phi}+\frac{1}{2}\alpha'(\phi)R+V'(\phi)=0,
\end{equation}
where the prime denotes the derivative with respect to the scalar field $\phi$.\\ The intrinsic Ricci scalar for a flat FRW brane is
\begin{equation}\label{courbure}
  R=6(\dot{H}+2H^{2}).
\end{equation}
\section{A Dynamical Systems Approach } \label{sec4}
In order to simplify the analysis of Eqs (\ref{freidman}), (\ref{Ray}) and (\ref{feild}), the method taken up is the dynamical systems study. In this section, we present the phase space of the non-minimally coupled scalar field in detail, exact solutions and their stability.\par
The first step in the implementation of the Dynamical System Approach (DSA) is the introduction of the general dimensionless variables
\begin{multline}\label{variables1}
  x_{1}\equiv \frac{1}{\sqrt{c} \kappa_{4}H}, \;\;\;\;x_{2}\equiv \frac{\sqrt{|\alpha(\phi)|}}{\sqrt{2c}H},\;\;\;\; x_{3}\equiv \frac{\dot{\alpha}(\phi)}{2cH^{3}},\\
  y\equiv \frac{\dot{\phi}}{\sqrt{12c}H^{2}},\;\;\;\; z\equiv \frac{\sqrt{V(\phi)}}{\sqrt{6c}H^{2}}.\;\;\;\;\;\;\;\;\;\;\;\;\;\
\end{multline}
The Friedmann constraint Eq. \eqref{freidman} with respect to the dimensionless variables \eqref{variables1} becomes
\begin{equation}\label{constraint}
  1= x_{1}^{2}(1- A x_{1}^{2})-x_{2}^{2}-y^{2}-z^{2}-x_{3}.
\end{equation}
The dynamical variables Eq. \eqref{variables1} are non-compact, i.e. their values do not have finite bounds as in \cite{ds1, ds2, ds3, ds4}. We will come back to this point in the conclusion.\\
The cosmological equations become equivalent to the following autonomous system:
{\small
\begin{subequations}
\begin{align}
  \frac{dx_{1}}{dN}  &= -x_{1} \frac{\dot{H}}{H^{2}} \label{dx1}\\
  \frac{dx_{2}}{dN} &= \frac{x_{3}}{2x_{2}}-x_{2}\frac{\dot{H}}{H^{2}} \label{dx2}\\
  \frac{dx_{3}}{dN} &= 6 \Gamma y^{2}-6 q x_{2} (3 y -6 q x_{2}+ r z)-3 (x_{3}+6 q^{2} x_{2}^{2})\frac{\dot{H}}{H^{2}} \label{dx3}\\
 \frac{dy}{dN} &=  -3y-6 q x_{2}-r z -\frac{\dot{H}}{H^{2}}(2y+3q x_{2}) \label{dy} \\
 \frac{dz}{dN} &= r y -2 z \frac{\dot{H}}{H^{2}}\label{dz}
\end{align}
\end{subequations}
}
where
{\small
\begin{widetext}
\begin{equation}\label{hpoint}
  \frac{\dot{H}}{H^{2}} = \frac{3}{2} \frac{x_{1}^{2}(1-A x_{1}^{2})+ x_{2}^{2} (-1+12 q^{2})+ y^{2}(1-2 \Gamma)-z^{2} -\frac{2}{3}x_{3}-1+2q x_{2}(3y+r z)}{1-x_{1}^{2}+(1-9 q^{2})x_{2}^{2}},
\end{equation}
\end{widetext}
}
the derivative is with respect to $N$ which is related to the scale factor $a$ by $N= \ln a$, and we define\\
\begin{multline}\label{const2}
 \qquad \quad \Gamma \equiv \alpha''(\phi),\quad r\equiv \frac{V'(\phi)}{\sqrt{2V(\phi)}H},\\
  q\equiv \frac{\alpha'(\phi)}{\sqrt{6 \alpha(\phi)}}\quad and\quad A\equiv \frac{c \kappa_{4}^{4}}{6}\lambda.\qquad\qquad
\end{multline}
The prime denotes the derivative with respect to the scalar field $\phi$.\par

From the first equation of the dynamical system \eqref{dx1}-\eqref{dz}, one can notice that the system has two invariant manifolds $x_{1} = 0$  and $\dot{H}/H^2= 0$. The most interesting, from a physical point of view, is the last one $\dot{H}/H^2= 0$.\par

The critical points  of any dynamical system can be  extracted by setting $dx_{i}/dN =0\; (i=1,2,...n)$, while their properties are determined by the eigenvalues $\mu_{i}$ of its Jacobian matrix, $J$, which is also called the stability matrix
\begin{equation}\label{Jacobian}
  J = \begin{pmatrix}
          \frac{\partial f_{1}}{\partial x_{1}}& \cdots & \frac{\partial f_{1}}{\partial x_{n}} \\
          \vdots & \ddots & \vdots\\
          \frac{\partial f_{n}}{\partial x_{1}} & \cdots & \frac{\partial f_{n}}{\partial x_{n}}
          \end{pmatrix},
\end{equation}
where $f_{i} \equiv (dx_{i}/dN)$.\\
The critical points are classified according to the sign of their eigenvalues by using linear stability method as:\\
 \begin{itemize}
   \item  Attractor critical point, If all eigenvalues have negative real parts. In this case the point would attract all nearby trajectories and is viewed as stable.\\
   \item  Repeller critical point, If all eigenvalues have positive real parts where trajectories are repelled from the fixed point and we speak in this situation of an unstable point.\\
   \item If there is mixture of both positive and negative real parts of eigenvalues, then the corresponding critical point is called a saddle. This point will attract nearby trajectories in some directions but repels them along others.\\
\end{itemize}
However, If at least one of the eigenvalues is zero, the linear stability theory fails to describe the stability of the critical point which is called non-hyperbolic. In this case other techniques have to be employed to study the stability properties, such as the Centre Manifold Theory (CMT) \cite{center,cm1,cm2,cm3}, the Lyapunov function method \cite{method, lyapunov1, lyapunov2} and Kosambi-Cartan-Chern theory \cite{Bohmer:2010re}.\\

\subsection{Example of $\alpha(\phi)$ and $V(\phi)$}
Since the dynamical system Eqs. \eqref{dx1}-\eqref{dz} is complicated to analyze in its full generality, we consider a particular case in order to illustrate our purpose. Concerning the scalar field potential, we choose an exponential function which has many implications in cosmological
inflation \cite{Copeland:1997et, Leon:2009rc}
\begin{equation}\label{potential}
  V= V_{0}\; e^{-b \kappa_{4}\phi},
\end{equation}
where $V_{0}$ corresponds to the maximum value of the potential and $b$ is a constant and we choose the following form of the coupling $\alpha(\phi)$ as
\begin{equation}\label{coupling}
  \alpha(\phi)= \alpha_{0} \; \phi^{2},
\end{equation}
where $\alpha_{0}$ is a constant parameter. If one chooses this monomial form of $\alpha(\phi)$, the set of phase space variables \eqref{variables1} reduces to a four dimensional by writing the variable $x_{3}$ as
\begin{equation}
  x_{3}=2\sqrt{6 \alpha_{0}}\; y\; x_{2}.
\end{equation}

{
In order to determine the energy density contribution of different components,  we define dimensionless density parameters in terms of the above introduced dynamical variables
{\small
\begin{multline}\label{omega}
 \Omega_\phi\equiv \frac{\kappa_{4}^{2} \rho_{\phi}}{6 H^{2}}=\frac{y^{2}+z^{2}}{x_{1}^{2}}, \Omega_\lambda\equiv \frac{\kappa_{4}^{2} \lambda}{6 H^{2}}=A x_{1}^{2}, \Omega_c\equiv  \kappa_{4}^{2} c H^{2}=\frac{1}{x_{1}^{2}},\\                                                                                                                                \Omega_\alpha\equiv \kappa_{4}^{2} \alpha_{0} (\frac{\phi}{2}+\frac{\dot{\phi}}{H})=\frac{x_{2}}{x_{1}^{2}}(x_{2}+2\sqrt{6\alpha_{0}}y),
\end{multline}
}which represent, respectively, the dimensionless energy densities of scalar field, effective cosmological constant, AdS/CFT correspondence effect and contribution due to the NMC corrections.\\
These dimensionless density parameters are related as
\begin{equation}\label{omega contraint}
  \Omega_\phi+\Omega_\alpha+\Omega_\lambda+\Omega_c=1.
\end{equation}
}

In the next subsections we will consider first two special values of $\alpha_{0}$, namely the minimal coupling for $\alpha_{0}=0$ and the conformal coupling for $\alpha_{0}=1/6$.

\subsubsection{Minimal coupling} \label{subsub1}
To illustrate our purpose, we begin by the simple case, namely the minimal one where $\alpha_{0}=0$. In that case the variable $x_{2}$ {and the energy density due to the NMC effect $\Omega_{\alpha}$ are} equal to zero . Using the Friedmann constraint Eq. (\ref{constraint}) and Eq.(\ref{potential}), the system \eqref{dx1}-\eqref{dz} reduces to the following autonomous two-dimensional system in terms of the dynamical variables
\begin{subequations}
\begin{align}
  \frac{dx_{1}}{dN}  &=- \frac{3 x_{1} y^2}{1-x_{1}^2}, \label{eq:dx1}\\
 \frac{dy}{dN} &=-3 y(1+ \frac{2 y^{2}}{1-x_{1}^{2}}) +\sqrt{3} b  \frac{-1-y^{2}+x_{1}^2(1-A x_{1}^{2})}{x_{1}}.\label{eq:dx2}
\end{align}
\end{subequations}

This nonlinear autonomous system has four critical points {\fontfamily{pzc}\selectfont A}$_{\pm}$ and {\fontfamily{pzc}\selectfont B}$_{\pm}$. Their properties are given in Table \ref{tab1}  and are summarized below.\\
\begin{table*}
  \centering
\begin{tabular}{|c|l|c|l|c|l|c|l|c|l|c}
  \hline
  Fixed points &\; \qquad $x_{1}$&$\quad y \quad$&$\quad z\quad$&Existence & \quad Eigenvalues& Stability& Physical State \\
  \hline  \hline
 {\fontfamily{pzc}\selectfont A}$_{\pm}$ &$\pm \sqrt{\frac{2}{1+\sqrt{1-4 A}}}$&$0$ &$\quad 0$&$A\leqslant \frac{1}{4}$&$\mu_{1}= 0$, $\mu_{2}=-3$ & Saddle for $A<0$ & \quad de Sitter\\
  &&&&&&Unstable for $0<A\leqslant \frac{1}{4}$&\quad universe\\  \hline
 {\fontfamily{pzc}\selectfont B}$_{\pm}$& $\pm\sqrt{\frac{1+\sqrt{1-4 A}}{2A}}$ &$0$&$\quad 0$&$0<A\leqslant \frac{1}{4}$&$\mu_{1}= 0$, $\mu_{2}=-3$&Unstable & \quad de Sitter  \\
  &&&&&&& \quad universe\\  \hline
\end{tabular}
  \caption{Coordinates of the critical points of the system \eqref{eq:dx1}-\eqref{eq:dx2}, with an exponential potential (\ref{potential})  and their properties.}
  \label{tab1}
\end{table*}
\begin{itemize}
  \item Critical points {\fontfamily{pzc}\selectfont A}$_{\pm}$ exist for $A\leq1/4$ (see Eq. \eqref{const2}) i.e. the effective cosmological constant on the brane satisfy  $\lambda\leq \lambda_{max}$ where, $\lambda_{max} \equiv 3/2c \kappa_{4}^{4}$. These points correspond to the case where the kinetic energy density of the scalar field and its potential energy density $V$ vanish (as the Hubble rate remains finite). This means that there is no dynamical motion of the the scalar field.\\
      The Freidmann equation of these fixed points writes
      \begin{equation}\label{FohA}
      H = \pm\sqrt{\frac{\Lambda_{-}}{3}},
      \end{equation}
 where $\Lambda_{-} = \frac{\lambda \kappa_{4}^{2}}{1-\sqrt{1-\frac{\lambda}{\lambda_{max}}}}$.\\
  Therefore, we conclude that the dynamic of the universe is governed by the effective cosmological constant. we notice also that the critical point  {\fontfamily{pzc}\selectfont A}$_{+}$ corresponds to an expanding de Sitter universe while {\fontfamily{pzc}\selectfont A}$_{-}$ represents a contracting one. \\
 \item Critical points {\fontfamily{pzc}\selectfont B}$_{\pm}$ exist only for a positive effective cosmological constant, in which the conformal anomaly is given by the condition $0<c\leq 3/ 2 \kappa_{4}^{4} \lambda$.\\
These points correspond to the solution:
       \begin{equation}\label{FohB}
         H = \pm\sqrt{\frac{\Lambda_{+}}{3}},
       \end{equation}
 where $\Lambda_{+} = \frac{\lambda \kappa_{4}^{2}}{1+\sqrt{1-\frac{\lambda}{\lambda_{max}}}}$.\\
Similar to the previous case, there is no dynamical motion of the the scalar field, and we have an expanding
de Sitter universe for the point {\fontfamily{pzc}\selectfont B}$_{+}$. The point {\fontfamily{pzc}\selectfont B}$_{-}$ represents a contracting de Sitter universe.
\end{itemize}

\begin{figure*}
  \centering
  \begin{tabular}{ccc}
    \subfloat[We have taken $A=-1$ and $b=-2$. \label{fig1a}]{\includegraphics[scale=0.45]{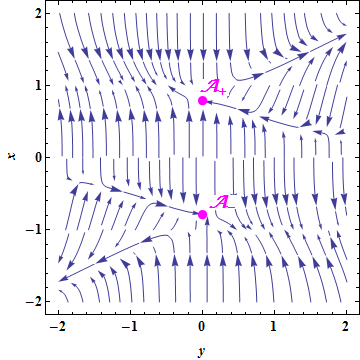}}& \qquad \qquad \qquad  &
     \subfloat[We have taken $A=1/8$ and $b=1$.\label{fig1b}]{\includegraphics[scale=0.45]{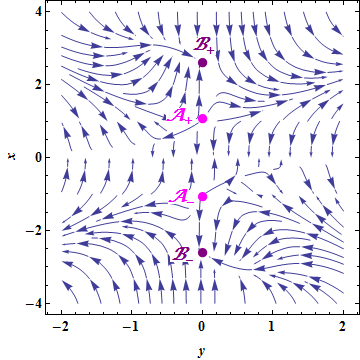}}
      \end{tabular}
    \caption[]{Phase plot (blue arrows) and critical points (colored dots) of the system \eqref{eq:dx1}-\eqref{eq:dx2}, for a minimal coupled scalar field with an exponential potential (\ref{potential}). It seems that point {\fontfamily{pzc}\selectfont B}$_{\pm}$ is saddle, but in fact it is  not, it is unstable from CMT point of view, see Appendix \ref{appendix}.}\label{figu1}
\end{figure*}

{
The eigenvalues corresponding to the critical points {\fontfamily{pzc}\selectfont A}$_{+}$ and {\fontfamily{pzc}\selectfont B}$_{+}$\footnote{We restrict our analysis to the critical points {\fontfamily{pzc}\selectfont A}$_{+}$ and {\fontfamily{pzc}\selectfont B}$_{+}$ since we are assuming an expanding universe, i.e. $H>0$.} are ($\mu_{1} =0$, $\mu_{2} =-3$)}.  We notice that these points are non hyperbolic. The stability properties of these points are obtained by applying the CMT to the 2D-system Eqs. \eqref{eq:dx1} and \eqref{eq:dx2} (the analysis details are given in the Appendix \ref{appendix}). Around the critical point {\fontfamily{pzc}\selectfont A}$_{+}$ the stability depends on the value of the constant $A$. For $A<0$ (i.e. $\lambda<0$), the critical point {\fontfamily{pzc}\selectfont A}$_{+}$ is saddle, whereas for $0<A\leq \frac{1}{4}$ both critical points  {\fontfamily{pzc}\selectfont A}$_{+}$ and  {\fontfamily{pzc}\selectfont B}$_{+}$ are unstable. Fig. \ref{figu1} shows the phase space and the position of the critical points of the system \eqref{eq:dx1}-\eqref{eq:dx2}.

We conclude that in the case of minimal coupling, the resulting Hubble rate can be considered as solutions at early times {with the energy density dominated by the effective cosmological constant and the AdS/CFT correspondence effect ($\Omega_{\lambda}+\Omega_{c}=1$)}. This means that in the past, each trajectory begins in a de Sitter state as the solution behaves like a cosmological constant of an arbitrary value for $A>0$.
\subsubsection{Conformal coupling}
 We now consider the case of a conformally coupled scalar field on the brane \cite{mariam, Faraoni:2000wk}, with conformal coupling $\alpha_{0} = \frac{1}{6}$, and a vanishing potential \footnote{ The Klein-Gordon equation \eqref{feild} is conformally invariant if $V=0$ or $V= \lambda \phi^{4}$ for the conformal coupling \cite{mariam, Faraoni:2000wk, Birrell, Wald}.}. In what follows, we present the results of our dynamical system \eqref{dx1}-\eqref{dz} for the conformal coupling. The system reduces to
\begin{subequations}
{\small
\begin{align}
    \frac{dx_{1}}{dN} \;&= -x_{1} \; g(x_{1}, x_{2}, y),\label{sx1} \\
    \frac{dx_{2}}{dN} \;&= y-x_{2}\;g(x_{1}, x_{2}, y),\label{sx2} \\
    \frac{dy}{dN}\;&= -3y - 2 x_{2} - (2 y+ x_{2})\; g(x_{1}, x_{2}, y) , \label{sx3}
 \end{align}
 }
 \end{subequations}
where
{\small
\begin{equation}\label{hpointmodif}
  g(x_{1}, x_{2}, y)=\frac{3 x_{1}^{2} (1 -A x_{1}^{2} )+ x_{2}^{2}+y^{2}+ 2 x_{2} y-3}{2 (1-x_{1}^{2})}.
\end{equation}
 }
In table \ref{tab:tab2}, we present the coordinates of each critical point and the results of their stability analysis by means of the signs of the real parts of the eigenvalues of the Jacobian matrix.\\
\begin{table*}
  \centering
\begin{tabular}{|c|l|c|l|c|l|c|l|c|l|}
  \hline
   Point & \; \qquad $x_{1}$ & \;\; $x_{2}$\;\;\; &\;\; $y$ \;\;\;  &Existence &\qquad \qquad \qquad Eigenvalues &Stability & Physical State  \\ \hline \hline
  {\fontfamily{pzc}\selectfont C}$_{\pm}$ &$\pm\sqrt{\frac{1-\sqrt{1-4 A}}{2A}}$ & $0$ &$\;\;\; 0$&$A\leqslant \frac{1}{4}$& $\mu_{1}= \frac{3(1-4A+ \sqrt{1-4A})}{2A}$, $\mu_{2}=-2$, $\mu_{3}=-1$ & Stable for $A<0$ &\quad de Sitter   \\
  &&&&&&Saddle for $0<A\leqslant \frac{1}{4}$& \quad universe\\  \hline
  {\fontfamily{pzc}\selectfont D}$_{\pm}$ &$\pm\sqrt{\frac{1+\sqrt{1-4 A}}{2A}}$ & $0$ &$\;\;\; 0$& $0<A\leqslant \frac{1}{4}$&$\mu_{1}= \frac{3(1-4A+ \sqrt{1-4A})}{2A}$, $\mu_{2}=-2$, $\mu_{3}=-1$& Saddle  & \quad de Sitter \\
  &&&&&&&\quad universe\\  \hline
\end{tabular}
  \caption{Coordinates of the critical points of the system \eqref{sx1}-\eqref{sx3}, and their properties.}
  \label{tab:tab2}
\end{table*}
\begin{itemize}
  \item Critical points {\fontfamily{pzc}\selectfont C}$_{\pm}$ exist for $A\leq 1/4$ and amount to assume that the solution is a de Sitter universe
  \begin{equation}\label{c+-}
     (H, \; \phi) = (\pm\sqrt{\frac{\Lambda_{-}}{3}}, 0).
  \end{equation}
  The stability of these two points depend on the value of the constant $A$ given by \eqref{const2}.\\
  For $A<0$ (i.e. $\lambda<0$), the critical points {\fontfamily{pzc}\selectfont C}$_{\pm}$ are stable since all eigenvalues are negative, whereas for $0<A\leq \frac{1}{4}$, the two critical points {\fontfamily{pzc}\selectfont C}$_{\pm}$ are saddle since one of the eigenvalues is positive while the others are negative.\\
  \item Critical points {\fontfamily{pzc}\selectfont D}$_{\pm}$ exist for $0<A\leq 1/4$ and represent also a de Sitter universe
  \begin{equation}\label{c+-}
     (H, \; \phi) = (\pm\sqrt{\frac{\Lambda_{+}}{3}}, 0).
  \end{equation}
  Finally, accordingly to their eigenvalues, the critical points {\fontfamily{pzc}\selectfont D}$_{\pm}$ are saddle points.
\end{itemize}
\begin{figure*}
\centering
  \begin{tabular}{ccc}
 \subfloat[Projection of perturbations along $x_{1}$-axis. \label{figa}]{\includegraphics[scale=0.5]{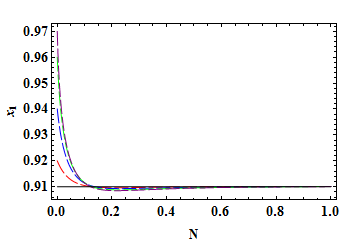}}& \qquad \qquad \qquad  &
\subfloat[Projection of perturbations along $x_{2}$-axis. \label{figb}]{\includegraphics[scale=0.5]{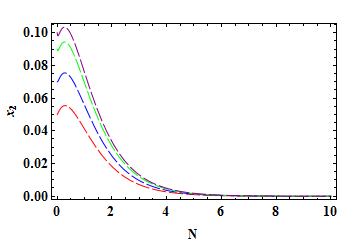}}\\
\subfloat[Projection of perturbations along $y$-axis. \label{figc}]{\includegraphics[scale=0.5]{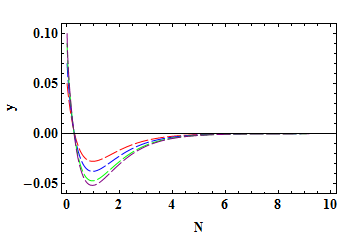}} & \qquad \qquad \qquad  &
\subfloat[Slow roll parameter, $\epsilon$. \label{epsilonplotc}]{\includegraphics[scale=0.5]{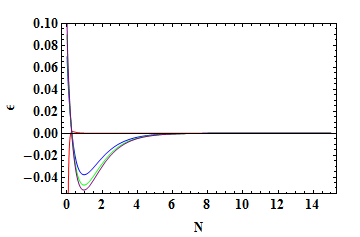}}
\end{tabular}
\caption[]{Projection of perturbations of {\fontfamily{pzc}\selectfont C}$_{+}$ along $x_{1}$, $x_{2}$, $y$ axis and slow roll parameter, $\epsilon$, vs the e-fold number, $N$, for $A=-1/4$.}\label{figure5}
\end{figure*}
\par

We notice that in the conformal coupling case, the critical point\footnote{We ignore {\fontfamily{pzc}\selectfont C}$_{-}$ since we are assuming only expanding universe.} {\fontfamily{pzc}\selectfont C}$_{+}$ is the future attractor if and only if $\lambda$ is negative while the critical point {\fontfamily{pzc}\selectfont D}$_{+}$ is always a saddle point meaning that it cannot be the past attractor. {Both {\fontfamily{pzc}\selectfont C}$_{+}$ and {\fontfamily{pzc}\selectfont D}$_{+}$ correspond to the energy density dominated by the effective cosmological constant and the AdS/CFT correspondence effect ($\Omega_{\lambda}+\Omega_{c}=1$).}\\
To confirm the stability of the critical point {\fontfamily{pzc}\selectfont C}$_{+}$, we perturb the solutions around this point in order to analyse numerically this property. In Figs. \ref{figa}, \ref{figb} and \ref{figc}, we plot the projection of perturbations of the system \eqref{sx1}-\eqref{sx3} along $x_{1}$-axis, $x_{2}$-axis and $y$-axis respectively with respect to $N$.\\
From these figures we notice that trajectories of the perturbed solutions approach the coordinates of {\fontfamily{pzc}\selectfont C}$_{+}$ for $A=-1/4$, i.e.  $x_{1} \simeq 0.91$, $x_{2} =0 $ and $ y= 0$ respectively as $N\rightarrow \infty$. From these behaviours, we can conclude that the critical point {\fontfamily{pzc}\selectfont C}$_{+}$ is an attractor solution which is in agreement with our analytical result.
Furthermore, this point corresponds to a slow roll parameter $\epsilon\equiv-\frac{\dot{H}}{H^{2}}=0$ which means that this point may sustain inflation. As we can notice from Fig. \ref{epsilonplotc}, the universe remains eternally in the inflation era even though we perturb it around this attractor solution.\par
\subsubsection{Non minimal coupling} \label{subsub2}
In what follows we will assume a positif non-minimal coupling constant $\alpha_{0}$ non equal to $0$ and $1/6$.\\
 In this subsection and due to the complexity of our system, we shall restrict our analysis to the case of $\lambda=0$, by choosing $\Lambda_{4} = \sqrt{ -6 \Lambda_{5}/\kappa^{2}}$ for $\Lambda_{5}<0$ in Eq. \eqref{lambda4}.\\
The set of the differentiable Eqs. \eqref{dx1}-\eqref{dz} reduces by considering the constraint Eq. (\ref{constraint}) to the following autonomous system
\begin{widetext}
{\small
\begin{align}
    \frac{dx_{1}}{dN} \;&= -x_{1} \; g(x_{1}, x_{2}, y),\label{sys1} \\
    \frac{dx_{2}}{dN} \;&=   \sqrt{6 \alpha_{0}}y-x_{2}\;g(x_{1}, x_{2}, y),\label{sys2} \\
    \frac{dy}{dN}\;&= -y(3+\frac{\sqrt{3}b}{x_{1}}(y+2 \sqrt{6 \alpha_{0}}x_{2})) -x_{2}(2\sqrt{6\alpha_{0}}+\frac{\sqrt{3} b}{x_{1}} x_{2})+
    \sqrt{3} b x_{1}- \frac{\sqrt{3}b}{x_{1}} - (2 y + \sqrt{6 \alpha_{0}} x_{2}) \;g(x_{1}, x_{2}, y), \label{sys4}
 \end{align}
 }
where
{\small
\begin{equation}\label{hpointmodif}
  g(x_{1}, x_{2}, y)= \frac{3 y^{2} \left( x_{1} (1-2 \alpha_{0})+\sqrt{6\alpha_{0}} b x_{2}\right)+3 x_{2} \sqrt{2\alpha_{0}}b \left(x_{2}^{2}-x_{1}^{2}\right)+2\sqrt{6\alpha_{0}}x_{2} y  (2 x_{1}+3\sqrt{6\alpha_{0}}b x_{2})+3 x_{2} (4 \alpha_{0} x_{1} x_{2}+\sqrt{2\alpha_{0}}b)}{x_{1}(1-x_{1}^{2}+x_{2}^{2}(1- 6 \alpha_{0}))}.
\end{equation}
 }
\end{widetext}

The fixed points of the system \eqref{sys1}-\eqref{sys4} are illustrated in table \ref{tab:tab1}.\\
\begin{table*}
  \centering
\begin{tabular}{|c|l|c|l|c|l|c|l|c|l|}
  \hline
   Point & $x_{1}$ & \;\;$x_{2}$ & \;$y$ \; & \;\;$z$&\;$\epsilon$\; &Existence & Stability & Description\\ \hline \hline
  {\fontfamily{pzc}\selectfont E} &$x_{1}$ & $ \frac{-\sqrt{2 \alpha_{0}}x_{1}+\sqrt{2\alpha_{0} x_{1}^{2}+b^{2}(-1+x_{1}^{2})}}{ b}$ &\;$0$&{\tiny $\sqrt{\frac{2 x_{1} \sqrt{2 \alpha_{0}(b^2 \left(x_{1}^2-1\right)+2 \alpha_{0}x_{1}^{2})}-4 \alpha_{0} x_{1}^2}{b^2}}$}&\;$0$&{\small $ x_{1}> 1$} \& {\small$b >0$}& Stable/ Saddle  & Potential\\
  &&&&&&&\qquad  Fig. \ref{fig2a}&domination  \\ \hline
  {\fontfamily{pzc}\selectfont F}  &$x_{1}$ &$ \frac{-\sqrt{2 \alpha_{0}}x_{1}-\sqrt{2\alpha_{0} x_{1}^{2}+b^{2}(-1+x_{1}^{2})}}{ b}$&\;$0$&{\tiny $\sqrt{\frac{-2 x_{1} \sqrt{2 \alpha_{0}(b^2 \left(x_{1}^2-1\right)+2 \alpha_{0}x_{1}^{2})}-4 \alpha_{0} x_{1}^2}{b^2}}$}&\;$0$&{\small $ x_{1}< -1$} \& {\small$b < 0$} &Stable/ Saddle & Potential\\
  &&&&&&&\qquad Fig. \ref{fig2b}&domination \\  \hline
\end{tabular}
  \caption{Critical lines, Stability, and the existence of the system \eqref{sys1}-\eqref{sys4} for an exponential potential \eqref{potential} and a non-minimal function \eqref{coupling} with $\lambda=0$.}
  \label{tab:tab1}
\end{table*}
\begin{figure*}
  \centering
  \begin{tabular}{ccc}
    \subfloat[Line {\fontfamily{pzc}\selectfont E}. \label{fig2a}]{\includegraphics[height=6cm]{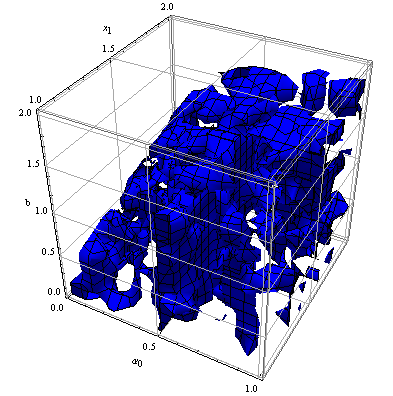}} & \qquad \qquad \qquad  &
    \subfloat[Line {\fontfamily{pzc}\selectfont F}. \label{fig2b}] {\includegraphics[height=6cm]{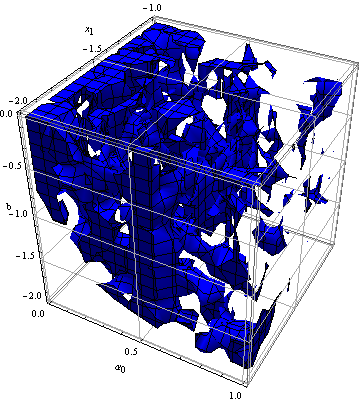}}
  \end{tabular}
    \caption[]{Blue region corresponds to the stable region of the critical lines {\fontfamily{pzc}\selectfont E} and {\fontfamily{pzc}\selectfont F}, while it is saddle otherwise.}\label{figure3}
\end{figure*}

\begin{itemize}
  \item The critical point {\fontfamily{pzc}\selectfont E} is formed of a continuous line of critical points, called a critical line or line of non-isolated equilibrium points \cite{nonisolated}. This critical line exists for an infinite number of critical points for all values of $x_{1}$ that verify the condition of existence $x_{1}>1$ and $b>0$. The dynamics of the universe for this critical line is dominated by the potential energy density, i.e. $V\neq0$ and $\dot{\phi}=0$ such that $\phi=\phi_{c}$. The Friedmann equation and the equation of motion of the scalar field of this critical line write respectively as
{\small
 \begin{equation}\label{fm}
   H^2 =  \frac{1}{4 c \kappa_{eff,c}^{2}} \left( 1\pm \sqrt{1-\frac{8 c\kappa_{eff,c}^{2}}{3} V(\phi_{c})}\right),
   \end{equation}
   }
   \begin{equation}\label{mm}
   V'(\phi_{c}) =  -6 \alpha'(\phi_{c}) H^2,
 \end{equation}
 where $V(\phi_{c}) < \frac{3}{8 c\kappa_{eff,c}^{2}}$.\\
The slow-roll parameter $\epsilon$ is equal to zero ($\epsilon =0$), which means that this critical line corresponds to inflation.
\begin{equation}\label{hy}
\end{equation}
\item The critical line {\fontfamily{pzc}\selectfont F} exists for $x_{1}<-1$.  We restrict our analysis to the critical line {\fontfamily{pzc}\selectfont E} since we assume an expanding universe ($H > 0$). Indeed, the critical line {\fontfamily{pzc}\selectfont F} does not correspond to an expanding universe due to the condition of existence for $x_{1} \propto H^{-1}$ (see table \ref{tab:tab1}).
\end{itemize}

\begin{figure*}
  \centering
  \begin{tabular}{ccc}
    \subfloat[Projection of perturbations along $x_{1}$-axis. \label{fig3a}]{\includegraphics[scale=0.5]{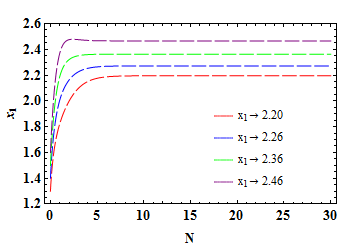}} & \qquad \qquad \qquad  &
    \subfloat[Projection of perturbations along $x_{2}$-axis. \label{fig3b}]{\includegraphics[scale=0.5]{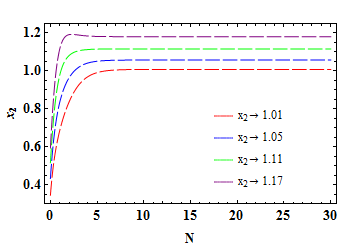}} \\
    \subfloat[Projection of perturbations along $y$-axis. \label{fig3c}]{\includegraphics[scale=0.5]{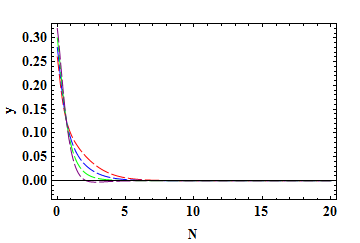}}& \qquad \qquad \qquad  &
    \subfloat[Projection of perturbations along $z$-axis. \label{fig3d}]{\includegraphics[scale=0.5]{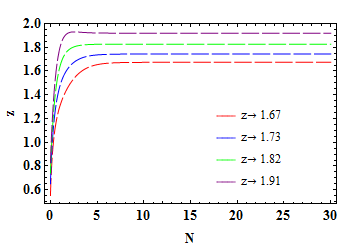}}\\
      \end{tabular}
    \caption[]{Projection of perturbations along $x_{1}$, $x_{2}$, $y$ axis for $\alpha_{0}= 0.2$ and $b=1$.}\label{figure4}
\end{figure*}

In order to discus the stability analytically, we use the linear theory. The stability of these lines is shown in\footnote{\label{note1} Note that the stability analysis of these critical lines depends on the value of the variable $x_{1}$, and the parameters of our model $\alpha_{0}$ and $b$ which makes the eigenvalues of the jacobian matrix very lengthy this is why we plot the stability region according to the signs of these eigenvalues.} Fig. \ref{figure3}. From both figures \ref{fig2a} and \ref{fig2b}, we conclude that the critical lines {\fontfamily{pzc}\selectfont E} and {\fontfamily{pzc}\selectfont F} are either stable or saddle. Consequently, the critical line {\fontfamily{pzc}\selectfont E} corresponds to a non-minimally coupled inflation attractor solution for a specific values of our model parameters $\alpha_{0}$ and $b$ in addition to the choice of the value of the dimensionless variable $x_{1}$. {This solution is dominated by the scalar field, the AdS/CFT correspondence effect and the NMC contribution($\Omega_{\phi}+\Omega_{c}+\Omega_{\alpha}=1$).}\\
   However the stability can also be found numerically by perturbing the system around the critical line. We plot in Fig. \ref{figure4} the projection plots on $x_{1}, x_{2}$, $y$ and $z$ separately for $\alpha_{0}= 0.2$ and $b=1$.
From Fig. \ref{fig3a} it seems that the trajectories are parallel to an horizontal axis, and that any perturbation of the system near $x_{1}$ makes it an arbitrary constant as $N\rightarrow\infty$.\\
We can also see from Fig. \ref{fig3b} and \ref{fig3d}, that for each value of $x_{1}$, the corresponding trajectories of $x_{2}$ and $z$ also approach the value $(\sqrt{b^2 \left(x_{1}^2-1\right)+2 \alpha_{0} x_{1}^2}-\sqrt{2 \alpha_{0} } x_{1}/b)$ and ${\small \sqrt{2 x_{1} \sqrt{2 \alpha_{0}(b^2 \left(x_{1}^2-1\right)+2 \alpha_{0}x_{1}^{2})}-4 \alpha_{0} x_{1}^2}/b}$ respectively as $N\rightarrow\infty$. Some numerical values of any perturbation near $x_{1}$, $x_{2}$ and $z$ are also shown in Fig. \ref{figure4}. For example for $x_{1}= 2.20$ the corresponding critical point coordinates $x_{2}$ and $z$ are $1.01$ and $1.67$ respectively as $N\rightarrow \infty$. \\
From Fig. \ref{fig3c}, we notice that trajectories of the perturbed solutions approach $y=0$ as $N\rightarrow \infty$.\\
From these behaviours it is evident that the system comes back to the critical point following the perturbation, which means that the critical line {\fontfamily{pzc}\selectfont E} is an attractor line for $\alpha_{0}= 0.2$ and $b=1$. These plots support strongly our analytical findings.\par

   In order to obtain a complete information about the structure of the phase space of the dynamical system \eqref{dx1}-\eqref{dz} it is necessary to investigate the dynamical behavior for $\alpha_{0}<0$. To this aim, we extend the previous study by including negative values of $\alpha_{0}$ in Eq. \eqref{coupling} to search for any possible attractor solutions. \\
 To keep the definition of the dimensionless variables as in \eqref{variables1}, one has to consider a non-minimal function as $\zeta (\phi)\equiv - \alpha (\phi)$, where $\alpha_{0}\equiv - \zeta_{0}$ and $\zeta_{0}$ is a positif constant. It deserves to be mentioned that the constraint equation Eq. \eqref{constraint} reads in this case
 \begin{equation}\label{constaintnmnegative}
  z^{2}= -1+x_{1}^{2}+x_{2}^{2}+2\sqrt{6 \alpha_{0}} y x_{2} - y^{2}.
 \end{equation}

  The set of differential Eqs. \eqref{dx1}-\eqref{dz} reduces to the following dynamical system
\begin{widetext}
  {\small
\begin{align}
    \frac{dx_{1}}{dN} \;&= -x_{1} \; g(x_{1}, x_{2}, y),\label{syst1} \\
    \frac{dx_{2}}{dN} \;&=   \sqrt{6 \zeta_{0}}y-x_{2}\;g(x_{1}, x_{2}, y),\label{syst2} \\
    \frac{dy}{dN}\;&= -y(3+\frac{\sqrt{3}b}{x_{1}}(y-2 \sqrt{6 \zeta_{0}}x_{2})) +x_{2}(2\sqrt{6\zeta_{0}}+\frac{\sqrt{3} b}{x_{1}} x_{2})+\sqrt{3} b x_{1}- \frac{\sqrt{3}b}{x_{1}}+(-2 y + \sqrt{6 \zeta_{0}} x_{2}) \;g(x_{1}, x_{2}, y), \label{syst3}
 \end{align}
 }
where
{\small
\begin{equation}\label{hpointmodif2}
  g(x_{1}, x_{2}, y)= \frac{3 y^{2} \left( x_{1} (1+2 \zeta_{0})-\sqrt{2\zeta_{0}} b x_{2}\right)+3 x_{2} \sqrt{2\zeta_{0}}b \left(x_{2}^{2}+x_{1}^{2}\right)+\sqrt{6\zeta_{0}}x_{2} y  (-4 x_{1}+6\sqrt{6\zeta_{0}}b x_{2})+3 x_{2} (4 \zeta_{0} x_{1} x_{2}-\sqrt{2\zeta_{0}}b)}{x_{1}(1-x_{1}^{2}-x_{2}^{2}(1+ 6 \alpha_{0}))}.
\end{equation}
 }
\end{widetext}

The system formed by the equations \eqref{syst1}-\eqref{syst3} has two critical lines. The coordinates of these critical lines with their qualitative behaviour are given in table \ref{tab:tab4}.\par
\begin{table*}[!hbtp]
  \centering
\begin{tabular}{|c|l|c|l|c|l|c|l|c|l|}
  \hline
   Point &\centering $x_{1}$ & \centering $x_{2}$ & \; $y$\; \; &\centering $z$&\;$\epsilon$ \;&Existence & Stability & Description\\ \hline \hline
  {\fontfamily{pzc}\selectfont G}&$x_{1}$ & $\frac{\sqrt{ -x_{1}^2 b^2+b^2+2 \zeta_{0} x_{1}^2}-\sqrt{2 \zeta_{0}} x_{1}}{b}$ &\;\;$0$&{\tiny $\sqrt{\frac{4 \zeta_{0} x_{1}^2-2 \sqrt{2 \zeta_{0} } x_{1} \sqrt{b^2-x_{1}^2 \left(b^2-2 \zeta_{0}\right)}}{b^2}}$}&\;$0$&{\small $ x_{1}> 1$} \&{\small$- \sqrt{\frac{2 \zeta x_{1}^{2}}{x_{1}^{2}-1}}<b <0$} &\;Stable & Potential d.\\ \hline
   {\fontfamily{pzc}\selectfont H}&$x_{1}$ &$ \frac{-\sqrt{ -x_{1}^2 b^2+b^2+2 \zeta_{0} x_{1}^2}-\sqrt{2 \zeta_{0}} x_{1}}{b}$&\;\;$0$&{\tiny $\sqrt{\frac{4 \zeta_{0} x_{1}^2+2 \sqrt{2 \zeta_{0} } x_{1} \sqrt{b^2-x_{1}^2 \left(b^2-2 \zeta_{0}\right)}}{b^2}}$}&\;$0$& {\small$b < 0$} &\;Saddle& Potential d. \\ \hline
\end{tabular}
  \caption{Critical lines, Stability, and the existence of the system Eqs. \eqref{syst1}-\eqref{syst3}
  .}
  \label{tab:tab4}
\end{table*}
\begin{figure}
  \centering
  \begin{tabular}{c}
    \subfloat[Line {\fontfamily{pzc}\selectfont G}. \label{figG}]{\includegraphics[height=5cm]{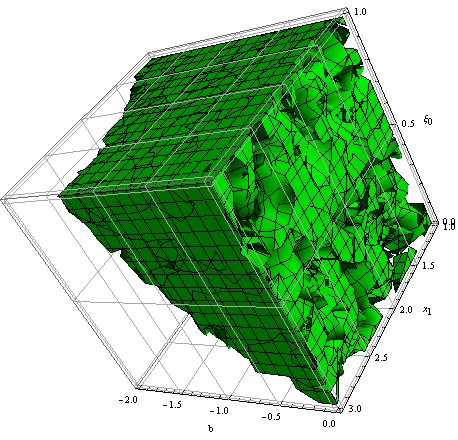}}\\
    \subfloat[Line {\fontfamily{pzc}\selectfont H}. \label{figH}] {\includegraphics[height=5cm]{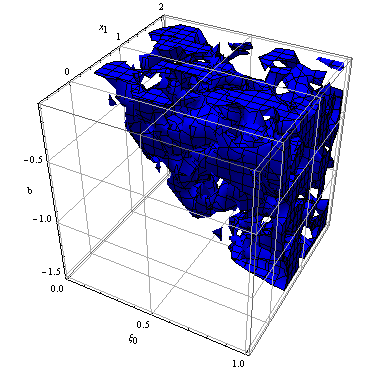}}
  \end{tabular}
    \caption[]{Blue (Green) region corresponds to the stable (saddle) region of the critical lines {\fontfamily{pzc}\selectfont G} and {\fontfamily{pzc}\selectfont H}.}\label{figureGH}
\end{figure}
\begin{itemize}
  \item For both critical lines {\fontfamily{pzc}\selectfont G} and {\fontfamily{pzc}\selectfont H} the dynamic of the universe is dominated by the potential energy density (as $\dot{\phi}$ vanishes while $V(\phi)\neq 0$) {in addition to the contribution of the AdS/CFT and the NMC effects ($\Omega_{\phi}+\Omega_{c}+\Omega_{\alpha}=1$)}, with the solution of the Hubble parameter writes as Eq. \eqref{fm}. The parameter $\epsilon$ evaluated at these critical lines is also equal to $0$ which means that these lines correspond to inflation.\\
\end{itemize}
\vspace*{1cm}
Examination of the stability conditions Fig. \ref{figureGH} indicates that the state can be stable (or saddle) during inflation. The critical line {\fontfamily{pzc}\selectfont G} is always saddle in the region of existence ($b<0$), while {\fontfamily{pzc}\selectfont H} is an attractor solution for $x_{1}>1$ and $- \sqrt{\frac{2 \zeta x_{1}^{2}}{x_{1}^{2}-1}}<b <0$.
\begin{figure}[h]
  \centering
  \begin{tabular}{c}
    \subfloat[Projection along $x_{1}$-axis. \label{fig5a}]{\includegraphics[scale=0.45]{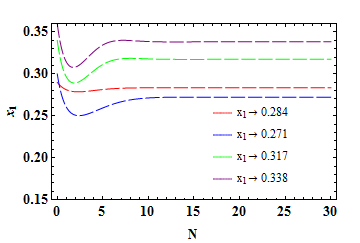}}\\
    \subfloat[Projection along $x_{2}$-axis. \label{fig5b}]{\includegraphics[scale=0.45]{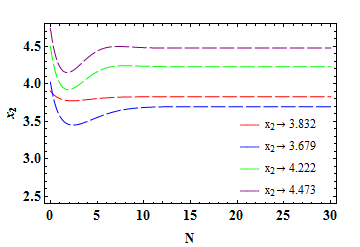}} \\
    \subfloat[Projection along $y$-axis. \label{fig5c}]{\includegraphics[scale=0.45]{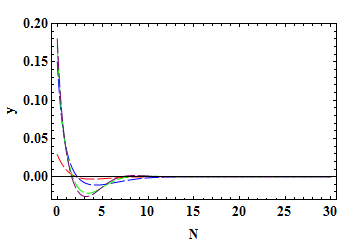}}\\
    \subfloat[Projection along $z$-axis. \label{fig5d}]{\includegraphics[scale=0.45]{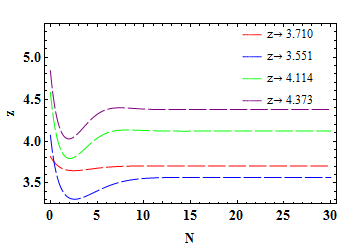}}\\
      \end{tabular}
    \caption[]{Projection of perturbations for $\zeta_{0}= 0.2$ and $b=-0.1$.}\label{figure5}
\end{figure}
To check the stability of the critical line {\fontfamily{pzc}\selectfont H} numerically, we perturb the solutions around the critical point. We again plot the projections plots on $x_{1}$, $x_{2}$, $y$ and $z$ separately for $\zeta_{0}= 0.2$ and $b=-0.1$. Like previous case, from Figs. \ref{fig5a}-\ref{fig5d}, it is clear that the critical line {\fontfamily{pzc}\selectfont H} is an attractor for $\zeta_{0}= 0.2$ and $b=-0.1$.\par
\section{Summary and conclusions} \label{conclusion}
{
The present work is devoted to the dynamical system analysis, in order to study the cosmological dynamics of NMC scalar field to the Ricci curvature within the AdS/CFT correspondence. We have considered an exponential potential $V= V_{0}\; exp(-b \kappa_{4}\phi)$ and a NMC function of the form $f(\phi)=\frac{1}{2\kappa_{4}^{2}} -\frac{1}{2} \alpha_{0} \phi^{2}$. The set of parameters characterizing our model are $(\alpha_0, \lambda, c, b)$, i.e. the coupling constant, the effective cosmological constant, the conformal anomaly coefficient and the free parameter of the potential, respectively.\par
}
The stability analysis of critical points is handled by using the linear stability analysis, the center manifold theory in the case of a non hyperbolic critical points and numerical techniques to support our results. The main results of this investigation can be summarized as follows
{
\begin{itemize}
  \item In the minimal coupling case, $\alpha_{0}=0$, the dynamics of the model is very simple. It consists of one non hyperbolic critical point {\fontfamily{pzc}\selectfont B}, which behaves as a past time attractor and a non hyperbolic critical point {\fontfamily{pzc}\selectfont A} corresponding to an unstable solution for $0<\lambda< 3/2c\kappa_4^4$ and to a saddle solution for $\lambda<0$. We have used the CMT to obtain sufficient conditions for their asymptotic stability. We also note that the solutions are described by a de Sitter state with the energy density dominated by the effective cosmological constant and the AdS/CFT correspondence effect ($\Omega_{\lambda}+\Omega_{c}=1$).
  \item Similar to the MC case, the solutions of the conformal coupling, $\alpha_{0}=1/6$ and $V=0$, are described by a de Sitter state with domination of the effective cosmological constant and AdS/CFT effect. We have identified two critical points {\fontfamily{pzc}\selectfont C} and {\fontfamily{pzc}\selectfont D}, the last one is always saddle while the point {\fontfamily{pzc}\selectfont C} is a future attractor point for $\lambda<0$ otherwise it behaves as a saddle point.
  \item In the non minimal coupling case, for $\alpha_{0}>0$ we have found one critical line, {\fontfamily{pzc}\selectfont E}, that corresponds to a future attractor de Sitter inflationary era for specific values of our model parameters $\alpha_{0}$ and $b$ (see Fig. \ref{fig2a}). For $\alpha_{0}<0$, we have found two critical lines, {\fontfamily{pzc}\selectfont G} and {\fontfamily{pzc}\selectfont H}. The critical line {\fontfamily{pzc}\selectfont H} represents a saddle line in its region of existence ($b<0$), while the critical line {\fontfamily{pzc}\selectfont G} is always a future attractor solution describing a de Sitter inflation scenario. For both $\alpha_{0}<0$ and $\alpha_{0}>0$ the dynamics of the universe is dominated by the potential energy density as well as the contribution of the AdS/CFT correspondence and the NMC effects ($\Omega_{\phi}+\Omega_{c}+\Omega_{\alpha}=1$).
\end{itemize}
}
Finally, one of the interesting results of including non-minimal coupling of the scalar field to the intrinsic curvature on the brane is the fact that we obtain a future attractor solution which corresponds to a scenario where the content of the universe is dominated by the exponential potential and a de Sitter inflationary era. {Another interesting conclusion consists in the fact that all the solutions are affected by the AdS/CFT correspondence, this can be seen through the contribution of the dimensionless energy density $\Omega_c$}. Despite the success of this study of non-minimal gravity within the AdS/CFT correspondence, the non compactness of our dynamical variables makes the analysis incomplete due to lack of the dynamical analysis at infinity of the phase space. Consequently, there could be missed critical points. This issue will be the subject of the next forthcoming paper.\par
\section*{Acknowledgement}
The authors would like to thank Mariam Bouhmadi-L\'{o}pez for many useful discussions and suggestions.\\
\appendix
\section{Centre Manifold Theory} \label{appendix}
In Sec. \ref{sec4}, we have mentioned that if the eigenvalues of the Jacobian matrix \eqref{Jacobian} has one eigenvalue with zero real part while the other one is negative, the critical point is called non-hyperbolic  and  the linear approach fails to determine the stability properties. Different methods can be employed to study the stability properties in this situation such as the  Lyapunov stability \cite{method, lyapunov1, lyapunov2}, centre manifold theory (CMT) \cite{center,cm1,cm2,cm3} and Kosambi-Cartan-Chern theory \cite{Bohmer:2010re}.\\
This Appendix is devoted to show how we get the stable conditions of the non-hyperbolic critical points {\fontfamily{pzc}\selectfont A}$_{+}$ and {\fontfamily{pzc}\selectfont B}$_{+}$ using the CMT.\\
In what follows, we present the detailed calculus to find the stable conditions of the critical point {\fontfamily{pzc}\selectfont A}$_{+}$.
To this purpose and in order to simplify the dynamical system Eqs. \eqref{eq:dx1}-\eqref{eq:dx2}, we define a new variable $\tau$ as $ d/d\tau = x_{1}(1-x_{1}^{2}) d/dN$.\\
We recall that for any dynamical system $\dot{x} = f (x)$, the new dynamical system $ \dot{x}= \chi(x) f(x)$, where $\chi(x)$ is a positive function, has the same critical points with the same stability properties. \\
For the critical point {\fontfamily{pzc}\selectfont A}$_{+}$, the function
$$
\left\{\begin{array}{lll}
\chi(x)&=x_{1}(1-x_{1}^{2}) \quad\text{is positive for}\; A<0\\
\chi(x)&= x_{1}(x_{1}^{2}-1)\quad\text{is positive for}\; A>0\\
\end{array}\right.
$$
Our dynamical system \eqref{eq:dx1}-\eqref{eq:dx2} becomes for $A<0$
\begin{subequations}
\begin{align}
  \frac{dx_{1}}{d\tau}  &=- 3 x_{1}^{2} y^2 \label{eq:dxm},\\
 \nonumber \frac{dy}{d\tau} &=-3 y x_{1}(1- x_{1}^{2}+2 y^{2}) \\
 &+\sqrt{3} b (-1-y^{2}+x_{1}^2(1-A x_{1}^{2}))(1-x_{1}^{2}).\label{eq:dxm}
\end{align}
\end{subequations}
 The first step is to consider a specific transformation: $X = x_{1}-k $ and $Y = y$ in order to move the critical point {\fontfamily{pzc}\selectfont A}$_{+}$($k,0$) to the origin of the phase space $(0,0)$, where $k= \sqrt{(1-\sqrt{1-4 A})/2A}$. We obtain the new dynamical system
\begin{subequations}
\begin{align}
  \frac{dX}{d\tau}  &=-3\; Y^2\; (k+X)^2, \label{eq:dxm1}\\
  \nonumber \frac{dY}{d\tau}&= 3 Y (k+X) \left(-1+(k+X)^2-2 Y^2\right)\\
  \nonumber &+\sqrt{3} b(k+X-1)(k+X+1)\\
  &\times \left((k+X)^2 \left(A(k+X)^2-1\right)+Y^2+1\right).\label{eq:dym1}
\end{align}
\end{subequations}
Our dynamical system has the required form, i.e. the fixed point sits at the origin $(0,0)$ and the system does not contain any linear term of $X$ in the first equation. We rewrite the above system as
{\small
\begin{equation}\label{cmA}
  \frac{d}{d\tau} \begin{pmatrix} X \\ Y \end{pmatrix}= \begin{pmatrix}  \mu_{1}&0\\ 0& \mu_{2} \end{pmatrix} . \begin{pmatrix} X\\Y \end{pmatrix}+ \begin{pmatrix} F(X,Y)\\ G(X,Y)\end{pmatrix},
\end{equation}}
where $\mu_{1}$ is the eigenvalue equal to zero, $\mu_{2}$ is a non-zero eigenvalue and, from Eqs. (\ref{eq:dxm1})-(\ref{eq:dym1}), the two functions $F$ and $G$ are
\begin{widetext}
\begin{align}
	F \left(X,Y\right)&= -3Y^2(k+X)^2,  \\
	\nonumber G\left(X,Y\right)&= 3 Y (k+X) \left(-2  Y^2+(k+X)^2-1\right) +Y^2 (k+X-1) (k+X+1)+3\sqrt{3}b X^2 k^2 \left(A \left(5 X^2-2\right)-2\right)\\
&+ \sqrt{3}b X^2\left(15 A k^4+20 A k^3 X+6 A k X^3-4 (A+1) k X+A X^4-(A+1) X^2+2\right).
\end{align}
\end{widetext}
and satisfy
\begin{align}\label{fg}
   F(0,0)= 0,  \qquad \nabla F(0,0)=0,\\
   G (0,0)=0,  \qquad \nabla G(0,0)=0.
\end{align}
The centre manifold (CM) suggests that  its geometrical space is tangent at $(0,0)$ to the eigenspace of the non zero eigenvalue $\mu_2$.
We may assume from the definition of the CM that $Y = H(X)$ with the following conditions:
\begin{equation*}
  H(0) = 0, \qquad \nabla  H(0) = 0.
\end{equation*}
In this coordinate, the dynamic of the CM, for $X$ sufficiently small, can be written as
\begin{equation}\label{cmy}
  \frac{dX}{d\tau}  = F\left( X,H(X)\right),
\end{equation}
 Assuming that $H(X)$ is of the form
\begin{equation}\label{HU}
  H(X) = a_{1} X^{2} + b_{1} X^{3}+{\cal{O}} (X^{4}),
\end{equation}
and using the Leibnitz rule $d Y/ d \tau = (d H/ d X) (d X/ d \tau )$, one obtains by combining the second row of Eq. \eqref{cmA} and Eq. \eqref{cmy}  the following equation
\begin{equation}\label{f}
  \frac{dH}{dX} F \left(X,H(X)\right) = \mu_{2} H(X) + G\left(X,H(X)\right)
\end{equation}
Comparing coefficients of equal order in $X$ of (\ref{f}), we find the coefficients $a_{1}$ and $b_{1}$
\begin{equation}
	a_{1}= \frac{b \left(-15 A k^4+6 (A+1) k^2-2\right)}{\sqrt{3} k \left(k^2-1\right)},
\end{equation}
\begin{equation}
 b_{1}= \frac{b \left(25 A k^6-(9 A+14) k^4+2 (A+4) k^2-2\right)}{\sqrt{3} k^2 \left(k^2-1\right)^2}.
\end{equation}
Since the system (\ref{eq:dxm1})-(\ref{eq:dym1}) has a CM, the evolution of this system  is given by
\begin{equation}\label{cme3}
\frac{dX}{d\tau} = -\frac{b^2\left(15 A k^4-6 (A+1) k^2+2\right)^2}{\left(k^2-1\right)^2}X^4+{\cal{O}}(X^5)
\end{equation}
We notice that the coefficient of the fourth order of $X$ is negative for $A<0$ and
consequently, around the fixed point {\fontfamily{pzc}\selectfont A}$_{+}$ the system is saddle in the centre manifold.\\We repeat the calculation in the case of $A>0$ for the fixed point {\fontfamily{pzc}\selectfont A}$_{+}$  and for the fixed point {\fontfamily{pzc}\selectfont B}$_{+}$ (since its existence is for $A>0$) we conclude that these points are unstable.


\begin{thebibliography}{20}
\bibitem{Randall:1999vf}
  L.~Randall and R.~Sundrum,
  Phys.\ Rev.\ Lett.\  {\bf 83} (1999) 4690
  [hep-th/9906064].

\bibitem{adsholo1}  G. ’t Hooft, Int. J. Mod. Phys. A 11,  (1996) 4623.

\bibitem{adsholo2} L. Susskind, J. Math. Phys. 36, (1995) 6377.

\bibitem{adsholo3} L. Susskind and E. Witten [hep-th/9805114].

\bibitem{adsholo4} E. Witten, Adv. Theor. Math. Phys. 2, (1998) 253.

 \bibitem{intro2}
  J.~D.~Brown and M.~Henneaux,
  Commun.\ Math.\ Phys.\  {\bf 104}  (1986) 207.

 \bibitem{intro3}  J.~M.~Maldacena,
  Int.\ J.\ Theor.\ Phys.\  {\bf 38} (1999) 1113
  [Adv.\ Theor.\ Math.\ Phys.\  {\bf 2}  (1998) 231]
  [hep-th/9711200].

 \bibitem{intro4} S.~S.~Gubser, I.~R.~Klebanov and A.~M.~Polyakov,
  Phys.\ Lett.\ B {\bf 428}  (1998) 105
  [hep-th/9802109].
	
	\bibitem{RSAds} Duff, M.J., and Liu, J.T.,
Phys. Rev. Lett., 85 (2000) 2052–2055,

  \bibitem{Dvali:2000hr}
  G.~R.~Dvali, G.~Gabadadze and M.~Porrati,
  Phys.\ Lett.\ B {\bf 485} (2000) 208
  [hep-th/0005016].

\bibitem{inducedper1} G. Dvali, G. Gabadadze and M. Porrati, Phys. Lett. B, (2000) 485, 208.

\bibitem{inducedper2} G. Dvali and G. Gabadadze, Phys. Rev. D, 63, (2001a), 065007.

\bibitem{inducedper3} G. Dvali, G. Gabadadze, M. Kolanovic and F. Nitti, Phys. Rev. D, 64, (2001b), 084004.

\bibitem{inducedper4} A. Lue, Phys. Rept., 423, (2006), 1.
\bibitem{Kofinas:2001es}
  G.~Kofinas,
  JHEP {\bf 0108} (2001) 034
  [hep-th/0108013].
  \bibitem{Deffayet:2000uy}
  C.~Deffayet,
  Phys.\ Lett.\ B {\bf 502} (2001) 199
  [hep-th/0010186].
\bibitem{Kiritsis:2002ca}
  E.~Kiritsis, N.~Tetradis and T.~N.~Tomaras,
  JHEP {\bf 0203} (2002) 019
  [hep-th/0202037].
\bibitem{Maeda:2003ar}
  K.~i.~Maeda, S.~Mizuno and T.~Torii,
  Phys.\ Rev.\ D {\bf 68} (2003) 024033
  [gr-qc/0303039].
\bibitem{Papantonopoulos:2004bm}
  E.~Papantonopoulos and V.~Zamarias,
  JCAP {\bf 0410} (2004) 001
  [gr-qc/0403090].
  \bibitem{Zhang:2004in}
  H.~s.~Zhang and R.~G.~Cai,
  JCAP {\bf 0408} (2004) 017
  [hep-th/0403234].
  \bibitem{BouhmadiLopez:2004ax}
  M.~Bouhmadi-Lopez, R.~Maartens and D.~Wands,
  Phys.\ Rev.\ D {\bf 70} (2004) 123519
  [hep-th/0407162].
  \bibitem{Maeda:2000mf}
  K.~i.~Maeda,
  Phys.\ Rev.\ D {\bf 64} (2001) 123525
  [astro-ph/0012313].

  \bibitem{GonzalezDiaz:2000ad}
  P.~F.~Gonzalez-Diaz,
  Phys.\ Lett.\ B {\bf 481} (2000) 353
  [hep-th/0002033].
  \bibitem{Majumdar:2001mm}
  A.~S.~Majumdar,
  Phys.\ Rev.\ D {\bf 64} (2001) 083503
  [astro-ph/0105518].
  \bibitem{Nunes:2002wz}
  N.~J.~Nunes and E.~J.~Copeland,
  Phys.\ Rev.\ D {\bf 66} (2002) 043524
  [astro-ph/0204115].
  \bibitem{Sami:2004ic}
  M.~Sami and N.~Dadhich,
  TSPU Bulletin {\bf 44N7} (2004) 25
  [hep-th/0405016].


 \bibitem{Futamase:1987ua}
  T.~Futamase and K.~i.~Maeda,
  Phys.\ Rev.\ D {\bf 39} (1989) 399.

\bibitem{Salopek:1988qh}
  D.~S.~Salopek, J.~R.~Bond and J.~M.~Bardeen,
  Phys.\ Rev.\ D {\bf 40} (1989) 1753.

\bibitem{Fakir:1990eg}
  R.~Fakir and W.~G.~Unruh,
  Phys.\ Rev.\ D {\bf 41} (1990) 1783.

\bibitem{Amendola:1990nn}
  L.~Amendola, M.~Litterio and F.~Occhionero,
  Int.\ J.\ Mod.\ Phys.\ A {\bf 5} (1990) 3861.

\bibitem{Kaiser:1994vs}
  D.~I.~Kaiser,
  Phys.\ Rev.\ D {\bf 52} (1995) 4295
  [astro-ph/9408044].

\bibitem{Bezrukov:2007ep}
  F.~L.~Bezrukov and M.~Shaposhnikov,
  Phys.\ Lett.\ B {\bf 659} (2008) 703
  [arXiv:0710.3755].

\bibitem{Bauer:2008zj}
  F.~Bauer and D.~A.~Demir,
  Phys.\ Lett.\ B {\bf 665} (2008) 222
  [arXiv:0803.2664].

\bibitem{Park:2008hz}
  S.~C.~Park and S.~Yamaguchi,
  JCAP {\bf 0808} (2008) 009
  [arXiv:0801.1722].

\bibitem{Linde:2011nh}
  A.~Linde, M.~Noorbala and A.~Westphal,
  JCAP {\bf 1103} (2011) 013
  [arXiv:1101.2652].

\bibitem{Kallosh:2013maa}
  R.~Kallosh and A.~Linde,
  JCAP {\bf 1310} (2013) 033
  [arXiv:1307.7938].

\bibitem{Kallosh:2013tua}
  R.~Kallosh, A.~Linde and D.~Roest,
  Phys.\ Rev.\ Lett.\  {\bf 112} (2014) no.1,  011303
  [arXiv:1310.3950].

\bibitem{Chiba:2014sva}
  T.~Chiba and K.~Kohri,
  PTEP {\bf 2015} (2015) no.2,  023E01
  [arXiv:1411.7104].

\bibitem{Boubekeur:2015xza}
  L.~Boubekeur, E.~Giusarma, O.~Mena and H.~Ramírez,
  Phys.\ Rev.\ D {\bf 91} (2015) 103004
  [arXiv:1502.05193].

\bibitem{Pieroni:2015cma}
  M.~Pieroni,
  JCAP {\bf 1602} (2016) no.02,  012
  [arXiv:1510.03691].

\bibitem{Salvio:2017xul}
  A.~Salvio,
  Eur.\ Phys.\ J.\ C {\bf 77} (2017) no.4,  267
  [arXiv:1703.08012].

  \bibitem{mariam}  M.~Bouhmadi-Lopez and D.~Wands,
  Phys.\ Rev.\ D {\bf 71} (2005) 024010
  [hep-th/0408061].

   \bibitem{Nozari:2012cy}
  K.~Nozari and N.~Rashidi,
  Phys.\ Rev.\ D {\bf 86} (2012) 043505
  [arXiv:1207.3966].
  \bibitem{Bogdanos:2006qw}
  C.~Bogdanos, A.~Dimitriadis and K.~Tamvakis,
  Phys.\ Rev.\ D {\bf 74} (2006) 045003
  [hep-th/0604182].
  \bibitem{Farakos:2006sr}
  K.~Farakos and P.~Pasipoularides,
  Phys.\ Rev.\ D {\bf 75} (2007) 024018
  [hep-th/0610010].
\bibitem{Accetta:1985du}
  F.~S.~Accetta, D.~J.~Zoller and M.~S.~Turner,
  Phys.\ Rev.\ D {\bf 31} (1985) 3046.
  \bibitem{deals}  V.~Faraoni,
  Phys.\ Rev.\ D {\bf 53} (1996) 6813
  [astro-ph/9602111].
  \bibitem{Quiros:2008hv}
  I.~Quiros, R.~Garcia-Salcedo, T.~Matos and C.~Moreno,
  Phys.\ Lett.\ B {\bf 670} (2009) 259
  [arXiv:0802.3362].
  \bibitem{Gonzalez:2008wa}
  T.~Gonzalez, T.~Matos, I.~Quiros and A.~Vazquez-Gonzalez,
  Phys.\ Lett.\ B {\bf 676} (2009) 161
  [arXiv:0812.1734].
\bibitem{Escobar:2012cq}
  D.~Escobar, C.~R.~Fadragas, G.~Leon and Y.~Leyva,
  Class.\ Quant.\ Grav.\  {\bf 29} (2012) 175006
  [gr-qc/1201.5672].

  \bibitem{ds1}  J.~A.~Leach, S.~Carloni and P.~K.~S.~Dunsby,
  Class.\ Quant.\ Grav.\  {\bf 23} (2006) 4915
  [gr-qc/0603012].
\bibitem{ds2} J.~D.~Barrow and S.~Hervik,
  Phys.\ Rev.\ D {\bf 74}  (2006) 124017
  [gr-qc/0610013].
\bibitem{ds3}  T.~Clifton and J.~D.~Barrow,
  Phys.\ Rev.\ D {\bf 72} (2005) 103005
  [gr-qc/0509059].

  \bibitem{Odintsov:2015wwp}
  S.~D.~Odintsov and V.~K.~Oikonomou,
  Phys.\ Rev.\ D {\bf 93} (2016) 023517
  [arXiv:1511.04559].

\bibitem{Odintsov:2017icc}
  S.~D.~Odintsov, V.~K.~Oikonomou and P.~V.~Tretyakov,
  Phys.\ Rev.\ D {\bf 96} (2017)  044022
  [arXiv:1707.08661].

    \bibitem{Odintsov:2017tbc}
      S.~D.~Odintsov and V.~K.~Oikonomou,
      Phys.\ Rev.\ D {\bf 96} (2017) 104049
  [arXiv:1711.02230].
  \bibitem{Dutta:2015jaq}
  J.~Dutta and H.~Zonunmawia,
  Eur.\ Phys.\ J.\ Plus {\bf 130} (2015) 221
  [gr-qc/1601.00283].

\bibitem{Escobar:2013js}
  D.~Escobar, C.~R.~Fadragas, G.~Leon and Y.~Leyva,
  Astrophys.\ Space Sci.\  {\bf 349} (2014) 575
  [gr-qc/1301.2570].

  \bibitem{Escobar:2011cz}
  D.~Escobar, C.~R.~Fadragas, G.~Leon and Y.~Leyva,
  Class.\ Quant.\ Grav.\  {\bf 29} (2012) 175005
  [gr-qc/arXiv:1110.1736].

  \bibitem{Zonunmawia:2018xvf}
  H.~Zonunmawia, W.~Khyllep, J.~Dutta and L.~Järv,
  Phys.\ Rev.\ D {\bf 98} (2018) 083532
  [gr-qc/1810.03816].
\bibitem{Sakstein:2015jca}
  J.~Sakstein and S.~Verner,
  Phys.\ Rev.\ D {\bf 92} (2015)  123005
  [gr-qc/1509.05679].

  \bibitem{Hrycyna:2015eta}
  O.~Hrycyna and M.~Szydłowski,
  JCAP {\bf 1511} (2015)  013
  [gr-qc/1506.03429].

  \bibitem{Bhattacharya:2015wlz}
  S.~Bhattacharya, P.~Mukherjee, A.~S.~Roy and A.~Saha,
  Eur.\ Phys.\ J.\ C {\bf 78} (2018)  201
  [gr-qc/1512.03902].

  \bibitem{jcap}  E.~Kiritsis,
  JCAP {\bf 0510} (2005) 014
  [hep-th/0504219].

\bibitem{boer}
  J. de Boer, E. Verlinde, and H. Verlinde, JHEP 08
 (2000) 003; E. Verlinde and H. Verlinde, JHEP. 05
(2000) 034; J. de Boer, Fortschr. Phys. 49  (2001) 339.

\bibitem{lidsey}  J.~E.~Lidsey and D.~Seery,
  Phys.\ Rev.\ D {\bf 73}  (2006) 023516
  [astro-ph/0511160].

  \bibitem{ds4}  S.~Carloni, P.~K.~S.~Dunsby, S.~Capozziello and A.~Troisi,
  Class.\ Quant.\ Grav.\  {\bf 22} (2005) 4839
  [gr-qc/0410046].

\bibitem{center} H. K. Khalil, Nonlinear Systems, 2nd edn (1996) (Englewood Cliffs, NJ: Prentice Hall), pp167-177.
\bibitem{cm1}  C.~G.~Boehmer, N.~Chan and R.~Lazkoz,
  Phys.\ Lett.\ B {\bf 714}  (2012) 11
  [arXiv:1111.6247].
\bibitem{cm2} J. Carr, Applications of Centre Manifold Theory, vol. 35 of Applied Mathematical Sciences. Springer US, New York, NY, 1981, 10.1007/978-1-4612-5929-9.
\bibitem{cm3}  A.~D.~Rendall,
  Gen.\ Rel.\ Grav.\  {\bf 34} (2002) 1277
  [gr-qc/0112040].
\bibitem{method} O. I. Bogoyavlensky, Methods in Qualitative Theory of Dynamical
Systems in Astrophysics and Gas Dynamic (Springer - Verlag,
Berlin 1985)
\bibitem{lyapunov1} S. Wiggins, Introduction to Applied Nonlinear Dynamical Systems
and Chaos. Springer, 1990.
\bibitem{lyapunov2}
 C.~G.~Boehmer and T.~Harko,
  J.\ Nonlin.\ Math.\ Phys.\  {\bf 17}  (2010) 503
  [arXiv:0902.1054].

\bibitem{Bohmer:2010re}
  C.~G.~Boehmer, T.~Harko and S.~V.~Sabau,
  Adv.\ Theor.\ Math.\ Phys.\  {\bf 16}  (2012) 1145
  [arXiv:1010.5464].

  \bibitem{Copeland:1997et}
  E.~J.~Copeland, A.~R.~Liddle and D.~Wands,
  Phys.\ Rev.\ D {\bf 57}  (1998) 4686
  [gr-qc/9711068].
\bibitem{Leon:2009rc}
  G.~Leon and E.~N.~Saridakis,
  JCAP {\bf 0911}  (2009) 006
  [arXiv:0909.3571].

  \bibitem{Faraoni:2000wk} V.~Faraoni,
  Phys.\ Rev.\ D {\bf 62} (2000) 023504
  [gr-qc/0002091].
  \bibitem{Birrell} Birrell, N.,  Davies, P. (1982). Quantum Fields in Curved Space (Cambridge Monographs on Mathematical Physics). Cambridge: Cambridge University Press. 
 \bibitem{Wald} R. M. Wald (1984). General relativity. Chicago, University of Chicago Press.

 \bibitem{nonisolated} A. A. Coley Dalhousie. Dynamical systems and cosmology Dalhousie University, Halifax, Canada.




\end{thebibliography}
\end{document}